\documentclass[submitting]{nst}

\usepackage{subfigure,dcolumn}
\usepackage[T2A,T1]{fontenc}
\usepackage[russian,english]{babel}

\usepackage{listings}
\begin{document}

\title{A unified classification-quantification framework for bubble-like nuclei within the extended quantum molecular dynamics model}
\thanks{Supported by the National Natural Science Foundation of China (NNSFC) (No. 12375123, No. 12475134, and No. U1832129), the Natural Science Foundation of Henan Province (No. 242300421048), the National Key Research and Development Program of China (No. 2022YFA1602404), and the Strategic Priority Research Program of Chinese Academy of Sciences (No. XDB34030000)}

\author{Ge Ren}
\affiliation{Centre of Theoretical Physics, College of Physics, Henan Normal University, Xinxiang 453007, China}
\affiliation{Shanghai Advanced Research Institute, Chinese Academy of Sciences, Shanghai, 201210, China}
\author{Chun-Wang Ma}
\email[Corresponding author, ]{machunwang@126.com}
\affiliation{Institute of Nuclear Science and Technology, Henan Academy of Sciences, Zhengzhou 450046, China}
\affiliation{Centre of Theoretical Physics, College of Physics, Henan Normal University, Xinxiang 453007, China}
\affiliation{Shanghai Research Center for Theoretical Nuclear Physics, NSFC and Fudan University, Shanghai, 200438, China}
\author{Xi-Guang Cao}
\email[Corresponding author, ]{caoxg@sari.ac.cn}
\affiliation{Shanghai Advanced Research Institute, Chinese Academy of Sciences, Shanghai, 201210, China}
\affiliation{Shanghai Institute of Applied Physics, Chinese Academy of Sciences, Shanghai, 201800, China}
\author{Kai-Xuan Cheng}
\affiliation{Centre of Theoretical Physics, College of Physics, Henan Normal University, Xinxiang 453007, China}
\author{Jie Pu}
\affiliation{Centre of Theoretical Physics, College of Physics, Henan Normal University, Xinxiang 453007, China}

\begin{abstract}
 A systematic study of relaxed low-energy cluster configurations for all nuclides listed
in the AME2020 database is performed within the extended quantum molecular dynamics (EQMD) framework, with frictional cooling enabling stable relaxation. A unified classification-quantification framework based on the dimensionless parameters $BHTU$ is established to characterize bubble-like nuclear morphologies. The factor $B$, determined from the number of inflection points in the radial density profile, categorizes nuclei into droplet ($B=0$), bubble ($B=1$), and toroidal bubble ($B=2$). The parameter $H$ defines the degree of central density depletion, while $T$ and $U$ characterize the relative surface thickness and the relative size of the internal low-density region, respectively. Light nuclei are predominantly droplet-like with $B=0$, $H=0$, $T=1$, $U=0$. Most medium-mass nuclei have $B=1$, consistent with previous studies, especially in the vicinity of $^{40}$Ca and the neutron-rich region, where nuclei show a pronounced central hollowing with large $H$ and $U$ values, identifying them as prime candidates for experimental searches for bubble structures. Toroidal bubble nuclei ($B=2$), emerging for $Z\approx25$ and prevalent in heavy systems, display a local density minimum at intermediate radius together with a shell-like low-density region. Furthermore, bubble structures are found to be widespread in the superheavy region, in agreement with earlier studies. This parameter scheme not only reveals the morphological richness of nuclei but also establishes a predictive framework for exploring exotic nuclear shapes, thereby opening new avenues for future theoretical and experimental investigations.
\end{abstract}


\maketitle
\nolinenumbers

\section{Introduction}

The development of radioactive and rare-isotope beams has greatly advanced the investigation of exotic-nucleus properties \cite{Ye2025,SBMa,Kondo2023,Fang2024} and of nuclear matter under extreme conditions \cite{ChenLW2024, Pujie2024}, along with the corresponding detection technologies \cite{Liao2024, SidaweiPRL2025}. However, understanding the structural diversity of atomic nuclei \cite{PJLPRL2023, Zhoubo.prc.N13.2025, khLi2025}, particularly the existence and characteristics of exotic shapes \cite{WuXY.NST2024}, remains a key challenge in nuclear physics. In the semiclassical liquid drop picture, the nucleus is typically viewed as a dense, nearly
spherical system whose central density approaches the saturation value of approximately 0.16 fm$ ^ {-3}$. However, studies suggest that, under specific conditions, especially in nuclei with large isospin asymmetry \cite{WanMT.NST2025, ShenLY.NST2025} and in excited states with high angular momentum\cite{B.Deyshan2020, Ichikawa2012}, nuclei can also exotic halo \cite{Tanihata1992, TANIHATA2013}, skin \cite{Abrahamyan2012, Adhikari2022, Duan.NST2024}, and even more exotic geometric structures, such as chain-like \cite{YeYL.PRL2020,YeYL.CP2023}, toroidal \cite{Wong1972,Ichikawa2012,CaoXG2019} and bubble configurations with significantly reduced central density \cite{WangYZ2011,LongWH2016}. This concept was first proposed by Wilson in 1946, who inferred the possible existence of bubble-like structures from analyses of low-lying excited states in shell-model nuclei \cite{Wilson}. Subsequently, researchers investigated the stability of bubble nuclei using droplet model and shell-correction methods, particularly their possible occurrence in medium-mass \cite{MG2009,LiZP2014,VChoudhary2020} and superheavy nuclei \cite{Bender1999,Decharge2003}.

Previous studies have shown that the defining characteristic of bubble nuclei is a significantly reduced nucleon density in their central region, a phenomenon that may be closely related to reduced occupancy of $s$ orbitals  \cite{AMutschler2017,VChoudhary2020}. Studies based on theoretical frameworks such as the self-consistent mean field (SCMF), the Hartree-Fock-Bogoliubov method (HFB) \cite{Decharge2003}, and the energy density functional approach (EDF) \cite{KHAN200837}, suggest that bubble structures may arise from a combination of shell effects, Coulomb repulsion, and an overall balance among nuclear interactions. Recent work on bubble nuclei has further stimulated both experimental and theoretical interest. For example, some experimental and theoretical studies have investigated whether $^{34}$Si exhibits a bubble structure through observables such as $\gamma$-ray  spectroscopy and spin-orbit splitting \cite{AMutschler2017, LiZP2013, ChenJ2024}. Comparable central depletion has also been predicted, especially near magic numbers \cite{Saxena2019}. These results are primarily based on static mean field theories. However, a recent study has suggested that dynamical models, such as the Extended Quantum Molecular Dynamics (EQMD) model, can also reveal the possible existence of bubble-like structures within cluster configurations \cite{RG2024,Natowitz2024}, which may be experimentally accessible through dynamical nonequilibrium states associated with collective excitations such as giant resonances.

Despite this progress, systematic calculations and a comprehensive mapping of bubble nuclei across the nuclear chart are still required. In addition, a unified  classification-quantification framework is needed to effectively describe the characteristics of their density distributions. Existing studies focus primarily on specific nuclides, such as Si \cite{ LiZP2013,FanXH2019,Grasso}, Ar \cite{KHAN200837, Campi1973, RG2024}, and local areas within the superheavy region \cite{Bender1999, Schuetrumpf2017}. A complete census of the nuclear chart outside these regions is lacking, in part due to challenges in experimental verification and technological limitations. Moreover, different theoretical frameworks exhibit inconsistencies in predicting bubble nuclei, especially when considering pairing correlations and deformation effects. These discrepancies highlight the need for a unified framework to systematically evaluate the possibility and physical properties of bubble nuclei. Unlike traditional mean-field methods that focus on smooth and nearly uniform density distributions, cluster formation naturally leads to local density variations, which may result in the formation of low-density regions between clusters \cite{Ebran2012, zhouboPRL2013, Wei2024, xufeilong2025}. In particular, centrosymmetric arrangements, such as tetrahedral or octahedral cluster configurations, can cause a significant reduction in the central density, thereby forming a ``bubble-like'' density distribution pattern with a central depression. This indicates that ``bubble-like'' features may be relatively common in cluster states. Therefore, identifying the distribution patterns of nuclei exhibiting such cluster-induced ``bubble-like'' features on the nuclide chart and establishing a systematic framework based on their characteristics are crucial for guiding experimental searches targeting these specific configurations.

Given its ability to simulate cluster dynamics and nonequilibrium configurations, EQMD offers a promising alternative for exploring bubble-like structures beyond the scope of conventional mean-field theories \cite{HeWB2014,RG2024}. The present work aims to systematically map the distribution of bubble-like structures in nuclear cluster configurations using the EQMD model and to develop a unified classification-quantification framework capable of revealing general structural patterns. This study is expected to deepen the understanding of the diversity of nuclear shapes and density distributions and to provide theoretical guidance for future experimental efforts in heavy-ion-collision studies targeting such exotic configurations.

\section{Methods}
\label{sec.M}

In studying exotic nuclear structures, the EQMD model has demonstrated significant advantages \cite{HeWB2014,B.Deyshan2020}, particularly for those involving clustering and non-uniform density distributions. This model is designed to handle nuclear configurations under nonequilibrium conditions and cluster-formation phenomena. First, the model modifies the treatment of zero-point kinetic energy to suppress unphysical quantum fluctuations that may distort the nuclear density distribution. Second, an effective Pauli potential is introduced to approximately account for the antisymmetrization effects of the many-body wave function, thereby better representing the Pauli exclusion principle and improving the description of short-range nuclear correlations. In addition, EQMD represents nucleons using Gaussian wave packets with complex width parameters, which dynamically evolve under the influence of the Hamiltonian, enabling precise tracking of spatial modulations and deformations in the nuclear density. Compared to the standard QMD model, which often encounters stability issues, the EQMD model adopts strict triple-loop calculations for the three-body term in Skyrme interactions and requires high-precision numerical solutions for the evolution equations. Although the computational cost is substantial, this treatment significantly enhances system stability and confines fluctuations within a minimal range. Finally, by integrating the friction cooling mechanism, unnecessary fluctuations are further reduced, enabling long-term stability of the system \cite{Maruyama}.

Unlike traditional mean-field approaches that assume static ground-state configurations, EQMD incorporates dynamic clusterization and frictional cooling, making it particularly well-suited for investigating unconventional nuclear shapes such as bubble-like structures. During the initialization phase, the introduction of the friction term serves to gradually reduce the energy of the system, stabilizing the initial configuration and favoring the emergence of clustered structures. This feature is particularly significant, as the formation of clusters within the nuclear medium typically leads to notable local density modulations, including increased densities within clustered regions and reduced densities in the spaces between them, which are essential for the emergence of bubble-like nuclear structures. Consequently, the EQMD model provides a robust framework for investigating bubble-like structures within clustered nuclear configurations, offering insights into their formation mechanisms under nonequilibrium conditions.

The equation of motion of the nucleus in EQMD is derived using the time-dependent variational principle. The total wave function of the system is expressed as a direct product of single-nucleon Gaussian wave packets 

\begin{equation} \label{eq:total_wave_function}
\begin{aligned}
\Psi = \prod_i \phi_i(\mathbf{r}),
\end{aligned}
\end{equation}
where $\phi_i(\mathbf{r})$ is the wave function of the $i$-th nucleon,written as
\begin{equation} \label{eq:ith_wave_function}
\begin{aligned}
\phi_i(\mathbf{r})=\left(\frac{v_i+v_i^*}{2 \pi}\right)^{3/4} \exp \left[-\frac{v_i}{2}\left(\mathbf{R}_i-\mathbf{r}\right)^2+\frac{i}{\hbar} \mathbf{P}_i \cdot \mathbf{r}\right].
\end{aligned}
\end{equation}
$\mathbf{R}_i$ and $\mathbf{P}_i$ are the coordinate and momentum centroids of the $i$-th wave packet. The complex Gaussian width $v_i$ is introduced as

\begin{equation} \label{eq:ith_wave_packet}
\begin{aligned}
v_i=\frac{1}{\lambda_i}+i \delta_i,
\end{aligned}
\end{equation}
in which $\lambda_i$ and $\delta_i$ are the real and imaginary parts of the width parameter, respectively. 

The system Hamiltonian comprises kinetic energy, center-of-mass correction $T_{c.m.}$, and effective interactions $H_{\text{int}}$, 

\begin{equation}
\label{eq:hamiltonian_en}
\begin{aligned}
H & = \langle \Psi | \sum_{i} \left( -\frac{\hbar^2}{2m} \nabla_i^2 - \widehat{T}_{c.m.} + \widehat{H}_{\text{int}} \right) | \Psi \rangle \\
& = \sum_{i} \left[ \frac{{\bf P}_i^2}{2m} + \frac{3\hbar^2 (1 + \lambda_i^2 \delta_i^2)}{4m\lambda_i} \right] - T_{c.m.} + H_{\text{int}}.
\end{aligned}
\end{equation}

$H_{\text{int}}$ contains Skyrme, Coulomb, symmetry, and Pauli terms characterized by the saturation density $\rho_0$ and constant parameters $\alpha=-124.3$ MeV, $\beta=70.5$ MeV, $\gamma=2.0$, $c_\text{S}=25$ MeV, $c_\text{P}=15$ MeV, and $f_0=1.0$,

\begin{equation}
\label{eq:int_potential_rewritten}
\begin{aligned}
H_{\text{int}} & = H_{\text{Skyrme}} + H_{\text{Coulomb}} + H_{\text{Symmetry}} + H_{\text{Pauli}}, \\
H_{\text{Skyrme}} & = \frac{\alpha}{2\rho_0} \int \rho^2({\bf r}) d^3r + \frac{\beta}{(\gamma+1)\rho_0^\gamma} \int \rho^{\gamma+1}({\bf r}) d^3r, \\
H_{\text{Symmetry}} & = \int \frac{c_\text{S}}{2} \frac{(\rho_p - \rho_n)^2}{\rho_0} d^3r, \\
H_{\text{Pauli}} & = \frac{c_\text{P}}{2} \sum_{i} (f_i - f_0)^\mu \theta(f_i - f_0).
\end{aligned}
\end{equation}

$f_i$ quantifies the overlap of the $i$-th nucleon with other nucleons of the same spin $S$ and isospin $T$,

\begin{equation}
f_i \equiv \sum_{j} \delta(S_i, S_j) \delta(T_i, T_j) |\langle \phi_i | \phi_j \rangle|^2.
\label{eq:overlap_en}
\end{equation}

Initially, nucleon positions are sampled uniformly within a sphere of radius $r=1.5A^{1/3}$, and momenta are subsequently drawn via Monte Carlo using the local Fermi gas approximation $k_F =\left( \frac{3\pi^2}{2} \rho \right)^{1/3}$, which is based on the density distribution of the sampled configuration. Relaxed low-energy cluster configurations can be obtained by solving the momentum dissipation equation from this initial random configuration.

\begin{equation}
\label{eq:friction}
\begin{aligned}
\dot{\mathbf{R}}_i = \frac{\partial H}{\partial \mathbf{P}_i} + \mu_{\mathbf{R}} \frac{\partial H}{\partial \mathbf{R}_i}, \quad
\dot{\mathbf{P}}_i = -\frac{\partial H}{\partial \mathbf{R}_i} + \mu_{\mathbf{P}} \frac{\partial H}{\partial \mathbf{P}_i}, \\
\frac{3\hbar}{4}\dot{\lambda}_i = -\frac{\partial H}{\partial \delta_i} + \mu_\lambda \frac{\partial H}{\partial \lambda_i}, \quad
\frac{3\hbar}{4}\dot{\delta}_i = \frac{\partial H}{\partial \lambda_i} + \mu_\delta \frac{\partial H}{\partial \delta_i},
\end{aligned}
\end{equation}
where $\mu_{\mathbf{R}}$, $\mu_{\mathbf{P}}$, $\mu_{\mathbf{\lambda}}$ and $\mu_{\mathbf{\delta}}$ are the damping coefficients. When these coefficients are
negative, the nucleus relaxes to a local energy minimum, and the wave packets stop evolving once Eq.~\eqref{eq:frictionstop} is satisfied.

\begin{equation}
\label{eq:frictionstop}
\begin{aligned}
\dot{\mathbf{R}}_i =0 \quad \dot{\mathbf{P}}_i =0 \\
\dot{\mathbf{\lambda}}_i =0 \quad \dot{\mathbf{\delta}}_i =0
\end{aligned}
\end{equation}

Applying the Wigner transform to Eq. (\ref{eq:ith_wave_function}), the phase space distribution function is obtained as

\begin{equation} \label{eq:Phase_space_function}
\begin{aligned}
 f(\mathbf{r},\mathbf{p})& = \frac{1}{(2 \pi \hbar)^{3/4}} \int \exp \left(\frac{i \mathbf{p} \xi}{\hbar}\right) \phi_i\left(\mathbf{r}^{-}\right) \phi_i^*\left(\mathbf{r}^{+}\right) d \xi \\
& =\frac{1}{(\pi \hbar)^3} \exp \left[-\frac{1+\lambda_i^2 \delta_i^2}{\lambda_i}\left(\mathbf{r}-\mathbf{R}_i\right)^2\right] \\
& \times \exp \left[-\frac{\lambda_i}{\hbar^2}\left(\mathbf{p}-\mathbf{P}_i\right)^2\right] \\
& \times\exp \left[\frac{2 \lambda_i \delta_i}{\hbar}\left(\mathbf{r}-\mathbf{R}_i\right) \cdot \left(\mathbf{p}-\mathbf{P}_i\right)\right], 
\end{aligned}
\end{equation}
where $\mathbf{r}^{\pm}=\mathbf{r} \pm \frac{\xi}{2}$ ($\xi$ is a small quantity). The coordinate-space density distribution of the nucleus is obtained by integrating $f(\mathbf{r}, \mathbf{p})$ over momentum space, namely,

\begin{equation}\label{rho_r}
\begin{aligned}
 \rho_i(\mathbf{r})&=\int f(\mathbf{r}, \mathbf{p}) d \mathbf{p}, \\ 
 \rho(\mathbf{r})&=\sum_i \rho_i(\mathbf{r}) =\sum_i \frac{1}{\left(\pi \lambda_i\right)^{3 / 2}} \exp \left[-\frac{\left(\mathbf{r}-\mathbf{R}_i\right)^2}{\lambda_i}\right] .
\end{aligned}
\end{equation}

\section{Results and Discussion}
\label{sec.rd}
\begin{figure}[!htb]
\includegraphics[width=\hsize] {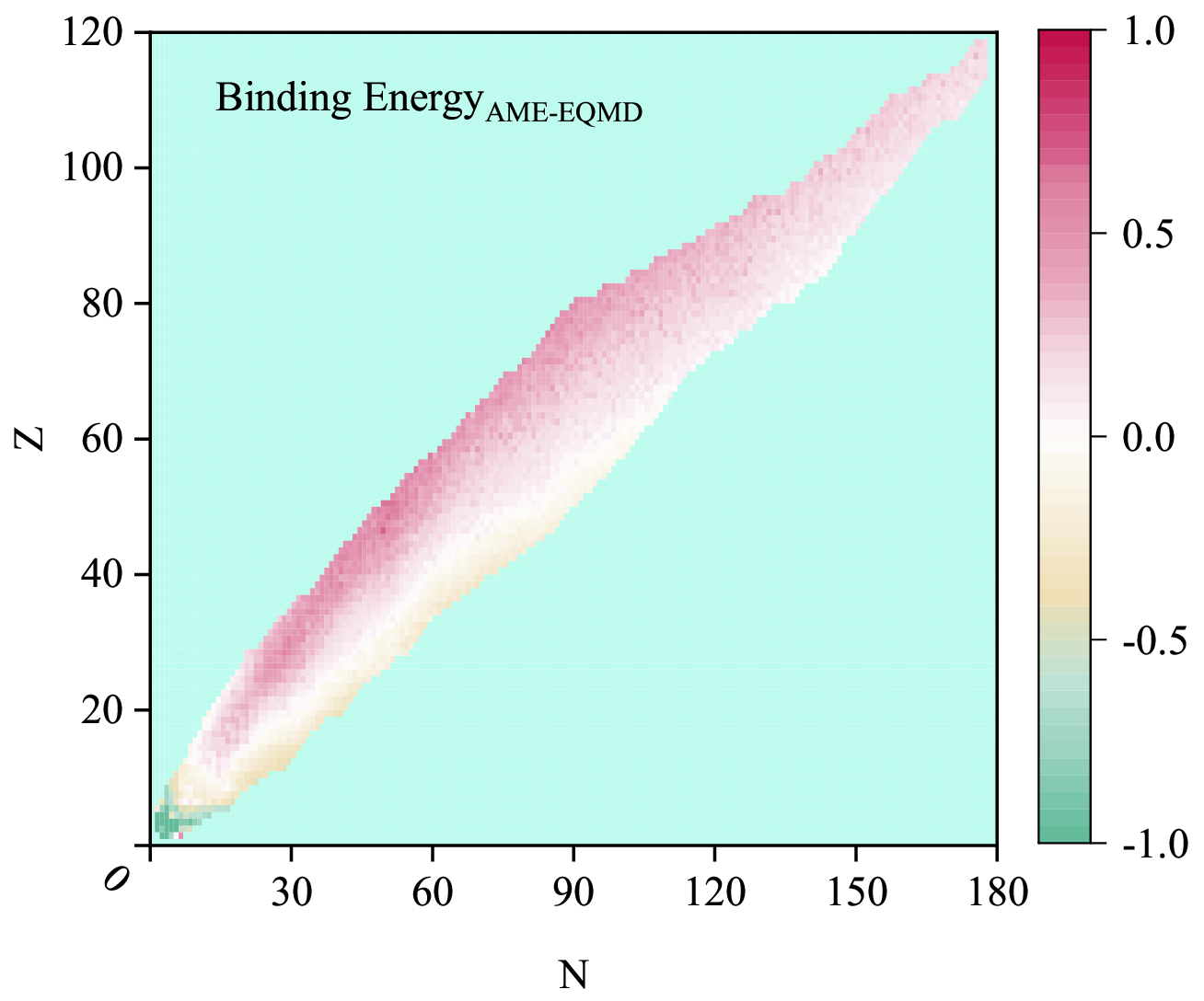}
\caption{(Color online) The difference in binding energy per nucleon, in MeV. Pink indicates that the EQMD binding energy is lower than the reference value (i.e. higher excitation energy).}
\label{fig:Ebind}
\end{figure}

Systematic calculations were performed for all nuclides in the AME2020 database using the EQMD model combined with the friction cooling method, producing relaxed low-energy cluster configurations. As described in Sec.~\ref{sec.M}, the improvements implemented in the EQMD model effectively counter the stability issues caused by feedback effects in the standard QMD density evolution, thereby minimizing the impact of fluctuations on energy and structure. To more strictly avoid the influence of statistical fluctuations, each nuclide was calculated at least three times, yielding an average standard deviation of the binding energy per nucleon of only 0.03 MeV for the full set of nuclides, indicating that the results can be considered convergent. Figure~\ref{fig:Ebind} depicts the difference in binding energy per nucleon between EQMD calculations and AME data. Significant deviations in the light nuclei region result from intense fluctuations in their intrinsic binding energy combined with high uncertainty from random sampling. For most other nuclei, the deviations cluster around zero, indicating relaxed states with low excitation energies close to the ground state. On the basis of their radial density distributions, these nuclides were classified into three distinct types. 

\begin{figure}[!htb]
\includegraphics[width=\hsize]{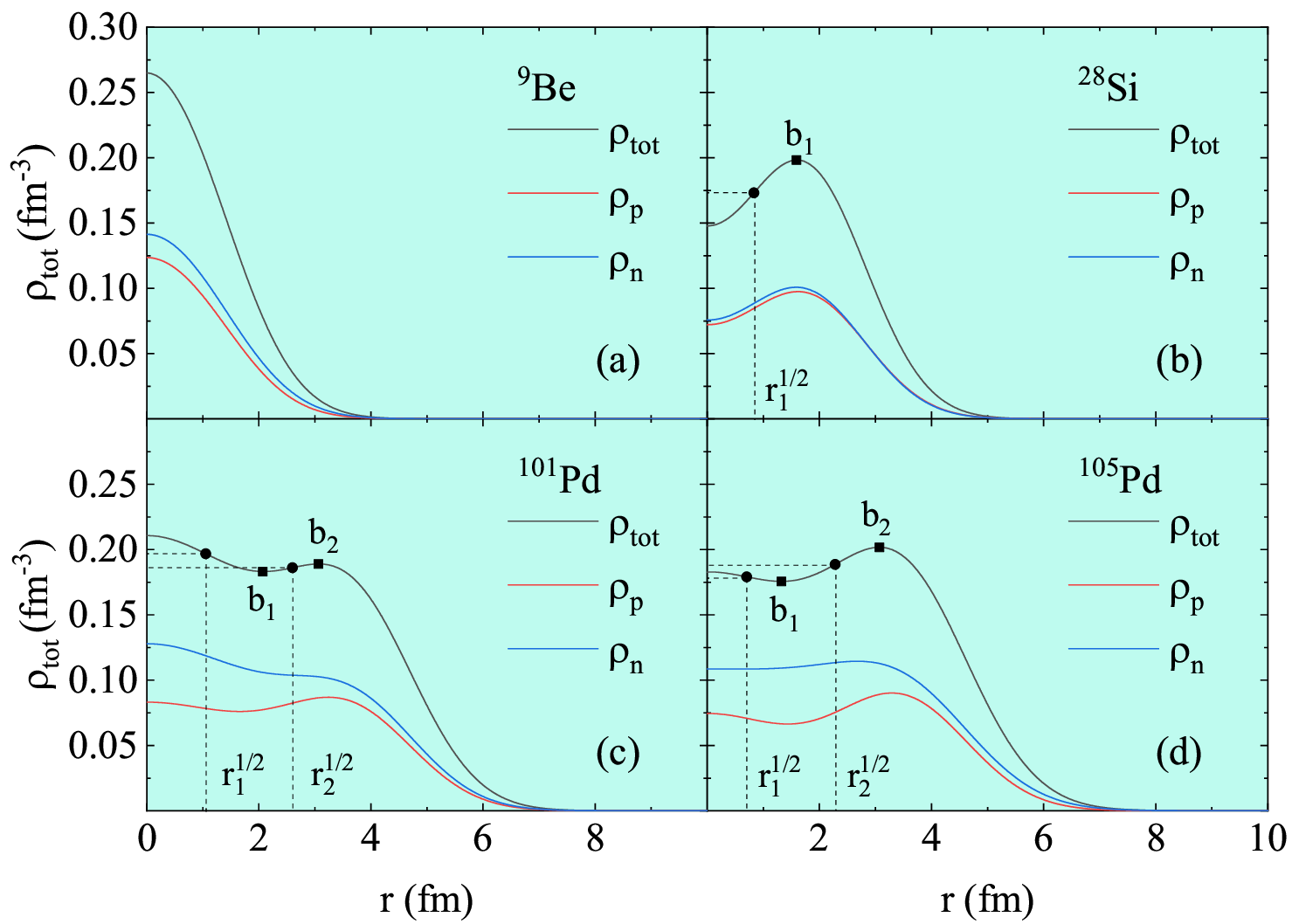}
\caption{(Color online) The total (black), proton (red) and neutron (blue) radial density distributions for droplet $^{9}$Be, bubble $^{28}$Si, and toroidal bubbles $^{101}$Pd and $^{105}$Pd. The $\text{b}_{n}$ are the $n$-th inflection points, and the $r_n^{1/2}$ correspond to the radius at which the radial density reaches the midpoint value between the $(n - 1)$-th and the $n$-th inflection points, starting from $r=0$, as parameters in Eqs.~(\ref{eq:T-Factor}) and (\ref{eq:U-Factor}}).
\label{fig:radial_density}
\end{figure}

As shown in Fig.~\ref{fig:radial_density}, the radial density of $^{9}$Be decreases monotonically with radius, consistent with a droplet-like wave-packet dispersion. The radial density of $^{28}$Si is consistent with the definition of a bubble nucleus, characterized by pronounced central density depletion. The two palladium isotopes, $^{101}$Pd and $^{105}$Pd, exhibit a similar phenomenon in which the density first decreases and then increases with increasing radius, indicative of a toroidal bubble structure in the mid-radius region of the nucleus. On this basis, the three representative morphologies are denoted droplet, bubble, and toroidal bubble. The toroidal bubble type can also be divided into two subtypes, one with a central density higher than the periphery, represented by $^{101}$Pd, and another with an  outer density higher than the center, exemplified by $^{105}$Pd.

\begin{figure}[!htb]
\includegraphics[width=\hsize]{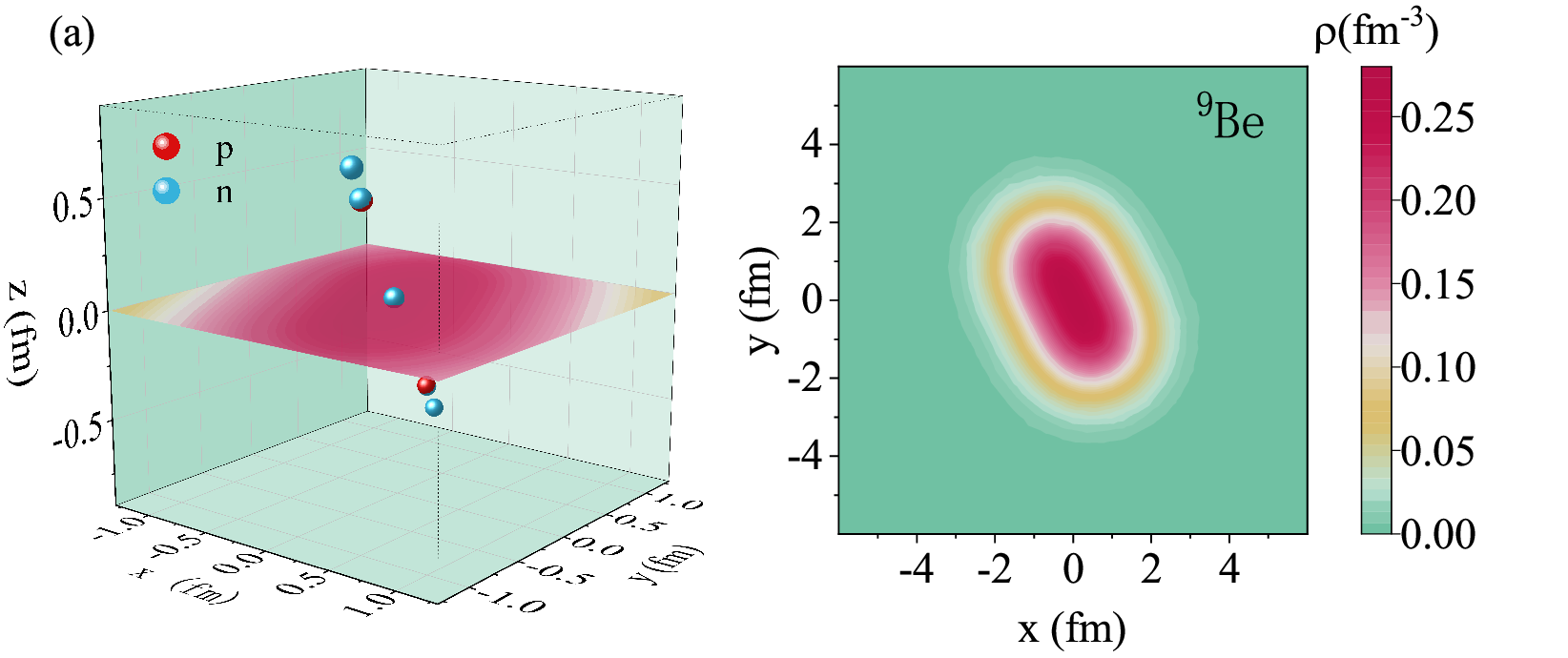}
\includegraphics[width=\hsize]{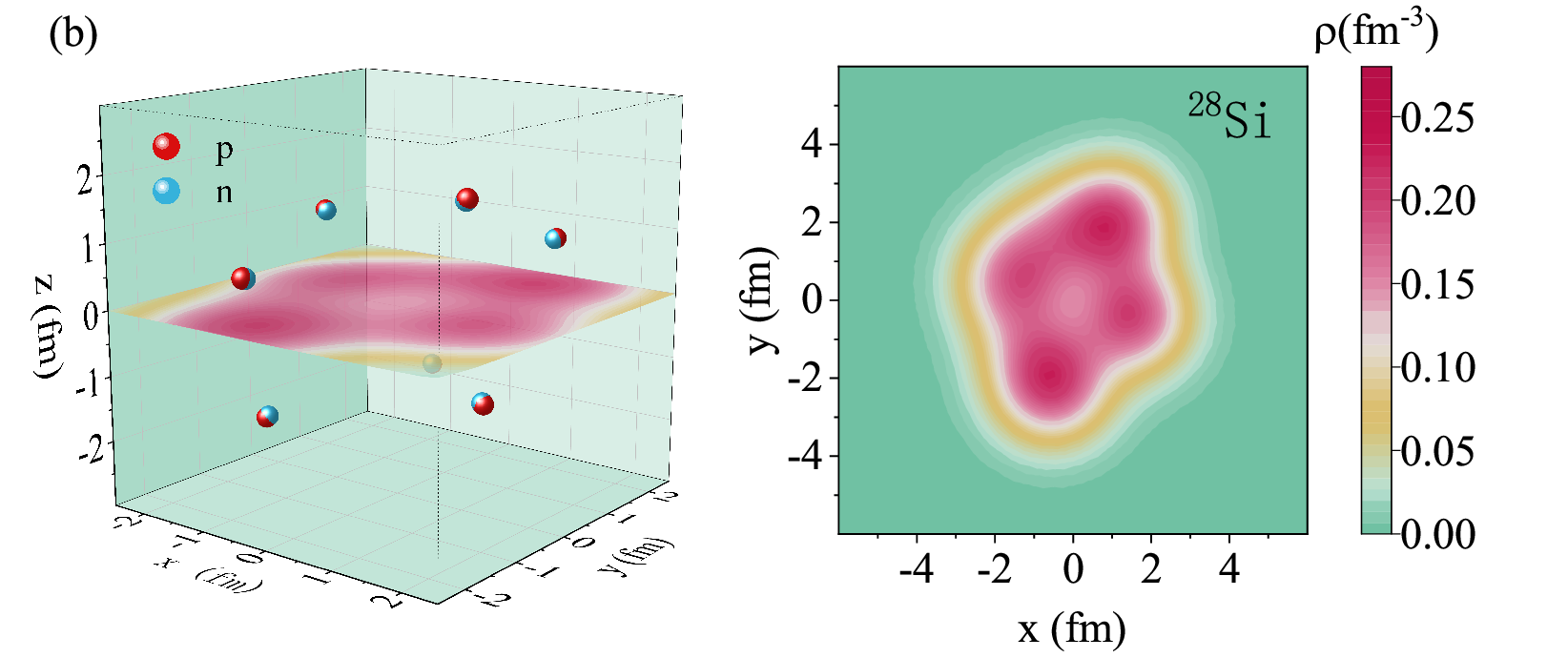}
\includegraphics[width=\hsize]{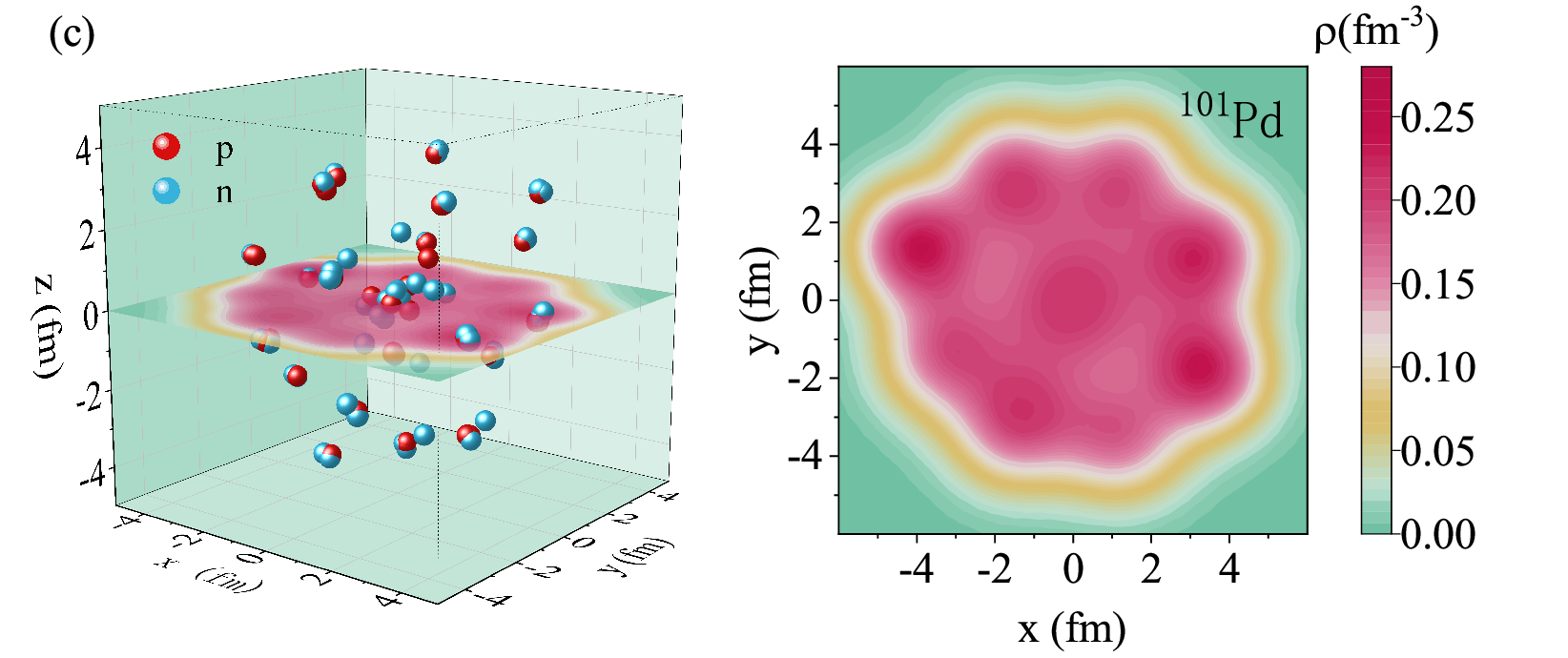}
\includegraphics[width=\hsize]{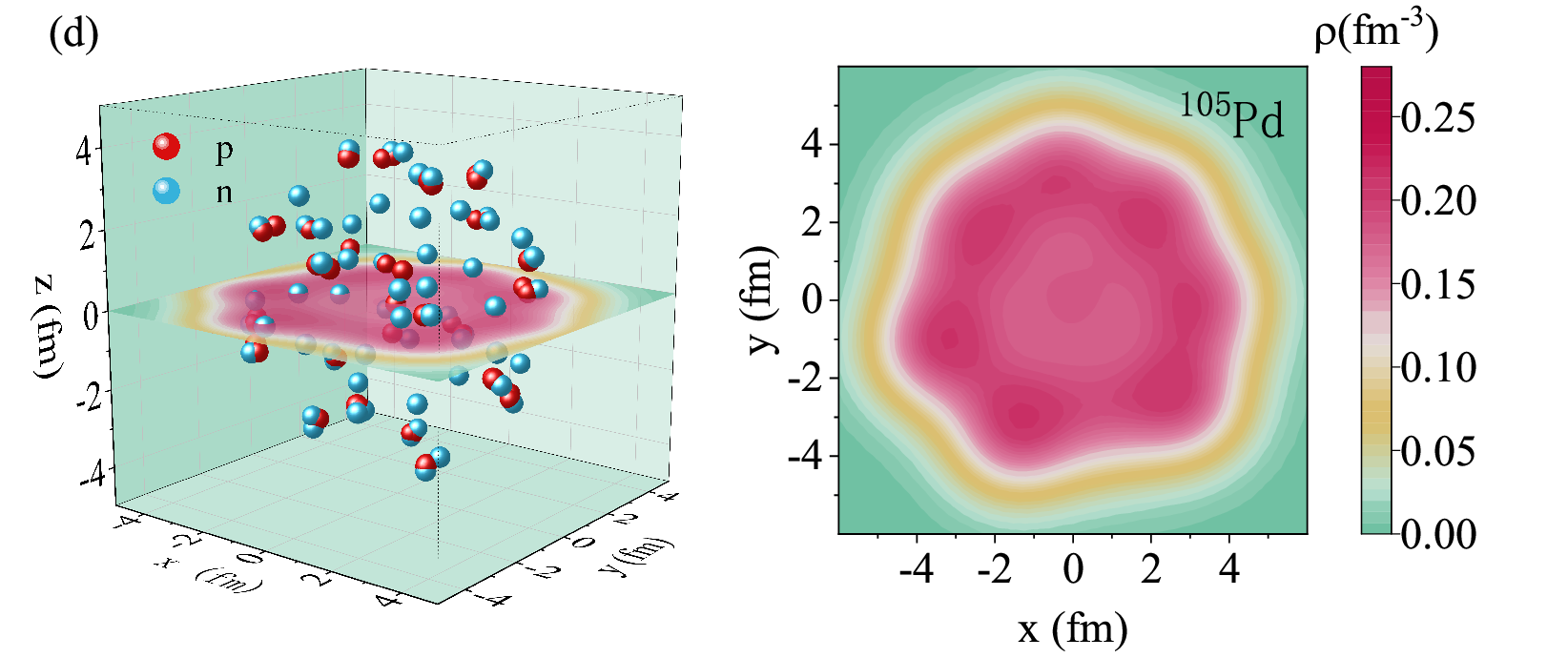}
\caption{(Color online) The centroids of nucleon wave packets in coordinate space (left) of $^{9}$Be, $^{28}$Si, $^{101}$Pd and $^{105}$Pd, and their corresponding density distributions  (right) at $z=0$.
}
\label{fig:density_distribution}
\end{figure}

Figure \ref{fig:density_distribution} presents the spatial distribution of the nucleon wave packet centers for nuclides $^{9}$Be, $^{28}$Si, $^{101}$Pd and $^{105}$Pd in the left panel, where the red spheres represent protons and the blue spheres denote neutrons, alongside their corresponding density profiles at $z=0$. The right panel visualizes the density distribution across the $x-y$ plane at $z=0$, using a shared color scale with the left panel to maintain consistency in the density representation. As shown in the left panel, under the frictional cooling mechanism of EQMD, neutrons and protons tend to cluster together, whereas excess neutrons in heavy nuclei tend to be delocalized and spread relatively uniformly across the nuclear volume. In detail, in Fig.~\ref{fig:density_distribution}~(a) $^{9}$Be exhibits a typical linear structure, with its density gradually decreasing from the center to the surface. This feature is consistent with the traditional wave-packet dispersion model. In light nuclei, the limited number of nucleons makes it necessary to have a sufficiently small internucleon spacing so that the strong interaction can provide enough binding energy to maintain a bound state. For cluster structures, when the distance between clusters is smaller than the spatial extension of the wave packets, significant overlap occurs between them. This overlap blurs the boundaries between clusters, making it difficult to form a distinct low-density region, and thus preventing the emergence of a pronounced central density depletion in the overall nuclear structure. In contrast, the seven $\alpha$ clusters of $^{28}$Si exhibit a significant quadrupole deformation and a certain degree of hexadecapole deformation \cite{GUPTA2023}. The sufficiently large inter-cluster spacing leads to a pronounced reduction in central density, which is a characteristic feature of a bubble-like nuclear structure. For the two palladium isotopes, $^{101}$Pd and $^{105}$Pd, there is clear clustering on the surface and a relatively high central nucleon density caused by wave packet overlap. This results in a radial density profile that initially decreases and then increases, giving rise to a toroidal bubble structure. Furthermore, $^{105}$Pd has a weaker degree of central clustering compared to $^{101}$Pd, effectively reducing the density in the nuclear center.

\begin{figure}[!htb]
\includegraphics[width=\hsize]{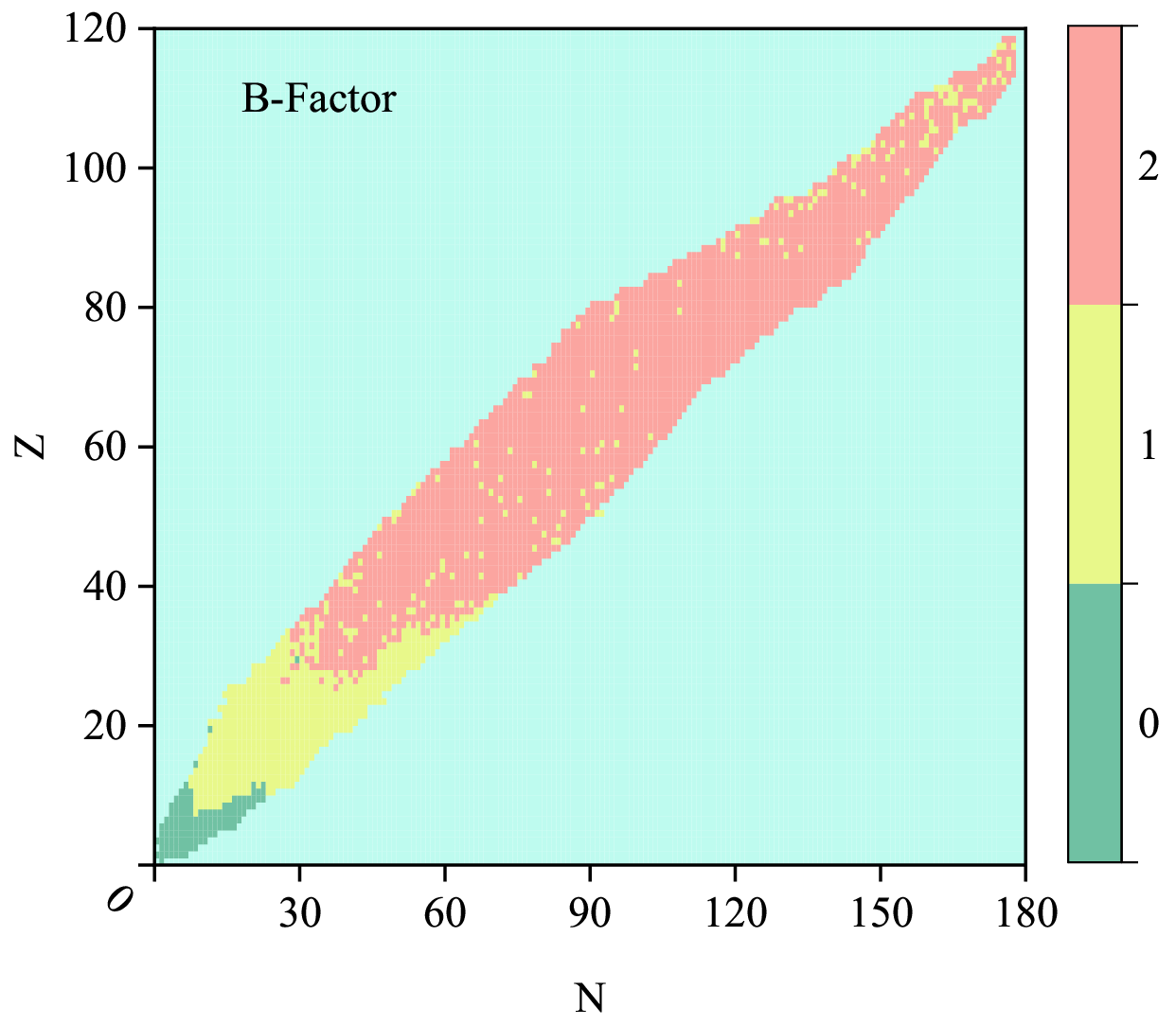}
\caption{(Color online) Three types of nuclides with different density characteristics reflected by the $B$ factor on the nuclear chart.  In this definition, $B=$ 0, 1, and 2  represent the droplet nucleus, bubble nucleus, and toroidal bubble nucleus, respectively.}
\label{fig:B_Factor}
\end{figure}

Instead of a simple classification, the primary objective is to understand the distribution patterns of nuclei with these different density characteristics on the nuclear chart. To achieve this, a factor $B$ was defined from the number of inflection points in the radial density profile as a criterion for distinguishing the three structural types: $B=0$, $B=1$ and $B=2$ correspond to droplets, bubbles and toroidal bubbles, respectively. resulting $B$ values for all nuclides are plotted on the nuclide chart in Fig. \ref{fig:B_Factor} as a nuclides chart. It is evident that, according to the EQMD calculations, droplet nuclei are mainly concentrated in the light-mass region, bubble structures mainly populate the intermediate-mass region, and toroidal bubble nuclei become increasingly prevalent in the region of $Z>28$.

It should be noted that toroidal bubble structures begin to emerge near the $\beta$-stable line at $Z \approx 25$, while the continuous distribution of bubble nuclei along the drip lines extends to heavier systems, reaching up to $Z = 38$ on the neutron drip line and forming a distinct $V$-shaped pattern. In line with previous studies, the number of bubble nuclei increases significantly in the superheavy region. These features can be understood as consequences of weak binding, which enhances the delocalization of surface nucleons and leads to exotic structures \cite{Ma.NST.2025, Zhoulong2024}. A similar $V$-shaped distribution also appears in the transition region between droplet and bubble nuclei around $Z=7-10$, because, in these light systems, the limited nuclear size and the increasing overlap of wave packets enhance the central density and thus suppress bubble formation near the drip lines.

Quantitative characterization of nuclear density distributions is essential for understanding the structural evolution across the nuclear chart, in addition to simply classifying nuclear structures. Following Ref.~\cite{MG2009}, the $H$ factor is defined to quantify the degree of central density depletion in relaxed low-energy cluster configurations as

\begin{equation}
\label{eq:Hfactor}
\begin{aligned}
H = \frac{\rho_{max}-\rho_{center}}{\rho_{max}},
\end{aligned}
\end{equation}
in which $\rho_{max}$ and $\rho_{center}$  denote the values of the maximum and central density, respectively. A larger $H$ therefore indicates a more pronounced reduction in central density relative to the periphery.

\begin{figure}[!htb]
\includegraphics[width=\hsize] {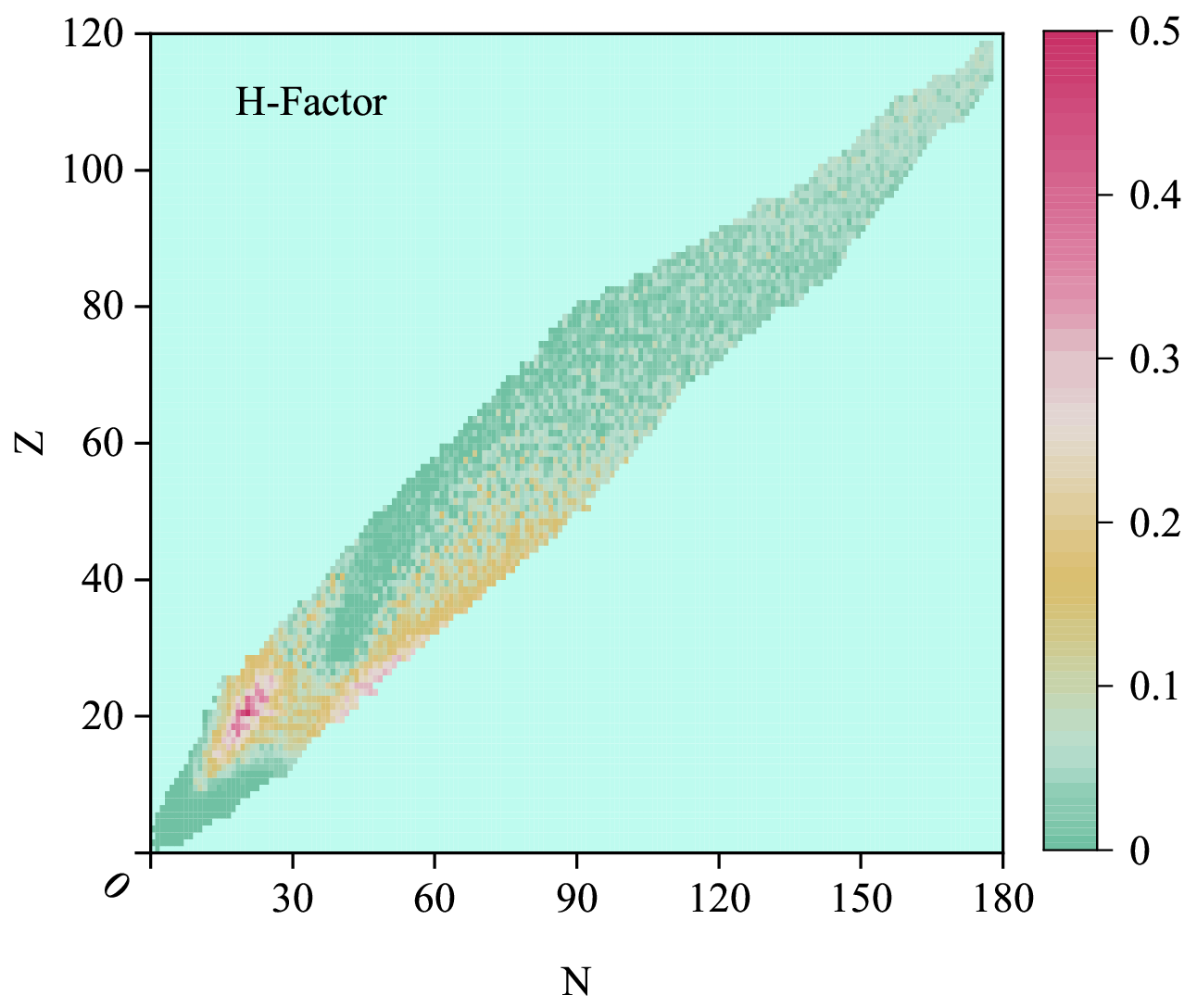}
\caption{(Color online) The $H$-factor for nuclides quantifies the degree of central hollowing in each nuclide. A higher $H$-factor indicates a greater reduction ratio in central density compared to the surrounding density.}
\label{fig:H_Factor}
\end{figure}

As shown in Fig.~\ref{fig:H_Factor}, for light-mass nuclei, droplet-like structures dominate, with $H$ values close to zero. In contrast, elevated $H$-factors are observed mainly in the region of intermediate-mass nuclei, in which bubble-like density profiles predominant among clustered configurations. The red coloration is especially prominent near $^{40}$Ca, the nuclide with the highest $H$, where the nearby nuclei exhibit relatively high stability, making them promising candidates for the experimental verification of bubble structures. 

Furthermore, the yellow and green colors observed in the toroidal bubble region of Fig.~\ref{fig:H_Factor} reflect distinct $H$ factor values that correspond to the two types of toroidal bubble illustrated in Figs.~\ref{fig:radial_density}(c) and (d), highlighting the capability of this metric to distinguish between different degrees of density depletion in this area. Combining the results shown in Fig.~\ref{fig:B_Factor} in the range $25<Z<50$, the difference in central density depletion near the proton and neutron drip lines reflects, to some extent, the difference in effective nuclear binding strengths between these two regions. However, for nuclei with $Z > 50$, as the number of nucleons increases, while a portion of nucleons form $\alpha$ clusters, the remaining unpaired nucleons tend to aggregate into other types of light cluster, such as H, He and Li, with different neutron numbers. The superposition of the wave packets of these non-$\alpha$ clusters' wave packets near the nuclear center led to an accumulation of density. Consequently, the formation of low-density regions in the center becomes increasingly suppressed, leading to a reduction in the $H$-factor values. As a result, green, which represents a higher central density ($\rho_{\text{center}}$), begins to dominate in this region.

To quantitatively capture the radial structure of the nuclear density more accurately, particularly the surface region and the central depletion, a pair of diagnostic parameters $T$ and $U$ are introduced to characterize, respectively, the thickness of the liquid surface and relative size of the low-density region. 

For bubble and toroidal bubble configurations, the dimensionless parameters $T$ and $U$ are defined as

\begin{align}
\label{eq:T-Factor}
T &= 
\begin{cases}
\displaystyle \frac{R_{\mathrm{rms}} - r_1^{1/2}}{R_{\mathrm{rms}}}, & B = 1, \\[1em]
\displaystyle \frac{R_{\mathrm{rms}} - r_2^{1/2}}{R_{\mathrm{rms}}}, & B = 2,
\end{cases} \\[1em]
\label{eq:U-Factor}
U &= 
\begin{cases}
\displaystyle \frac{r_1^{1/2}}{R_{\mathrm{rms}}}, & B = 1, \\[1em]
\displaystyle \frac{r_2^{1/2} - r_1^{1/2}}{R_{\mathrm{rms}}}, & B = 2,
\end{cases}
\end{align}
where $R_{\mathrm{rms}}$ is the root-mean-square radius of the nucleus, and $r_n^{1/2}$ denotes the radius at which the radial density profile reaches the midpoint value between the $(n-1)$-th and the $n$-th inflection points, starting from $r=0$, which can be more readily understood from the dots and dotted-line annotations in Fig~\ref{fig:radial_density}. Equivalently, in a more intuitive form,

\begin{align}
\label{eq:TU-Factors2}
T &= \frac{L_{\mathrm{surf}}^{1/2}}{R_{\mathrm{rms}}}, \\
U &= \frac{L_{\mathrm{bubble}}^{1/2}}{R_{\mathrm{rms}}},
\end{align}
in which $L_\mathrm{surf}$ and $L_\mathrm{bubble}$ correspond to the radial extents of the surface liquid-film and low-density bubble regions, respectively, each defined as the distance between the two radii where the density equals half of its local maximum. For a droplet configuration, one has $T = 1$ and $U = 0$ by this definition.

\begin{figure}[!htb]
\includegraphics[width=\hsize] {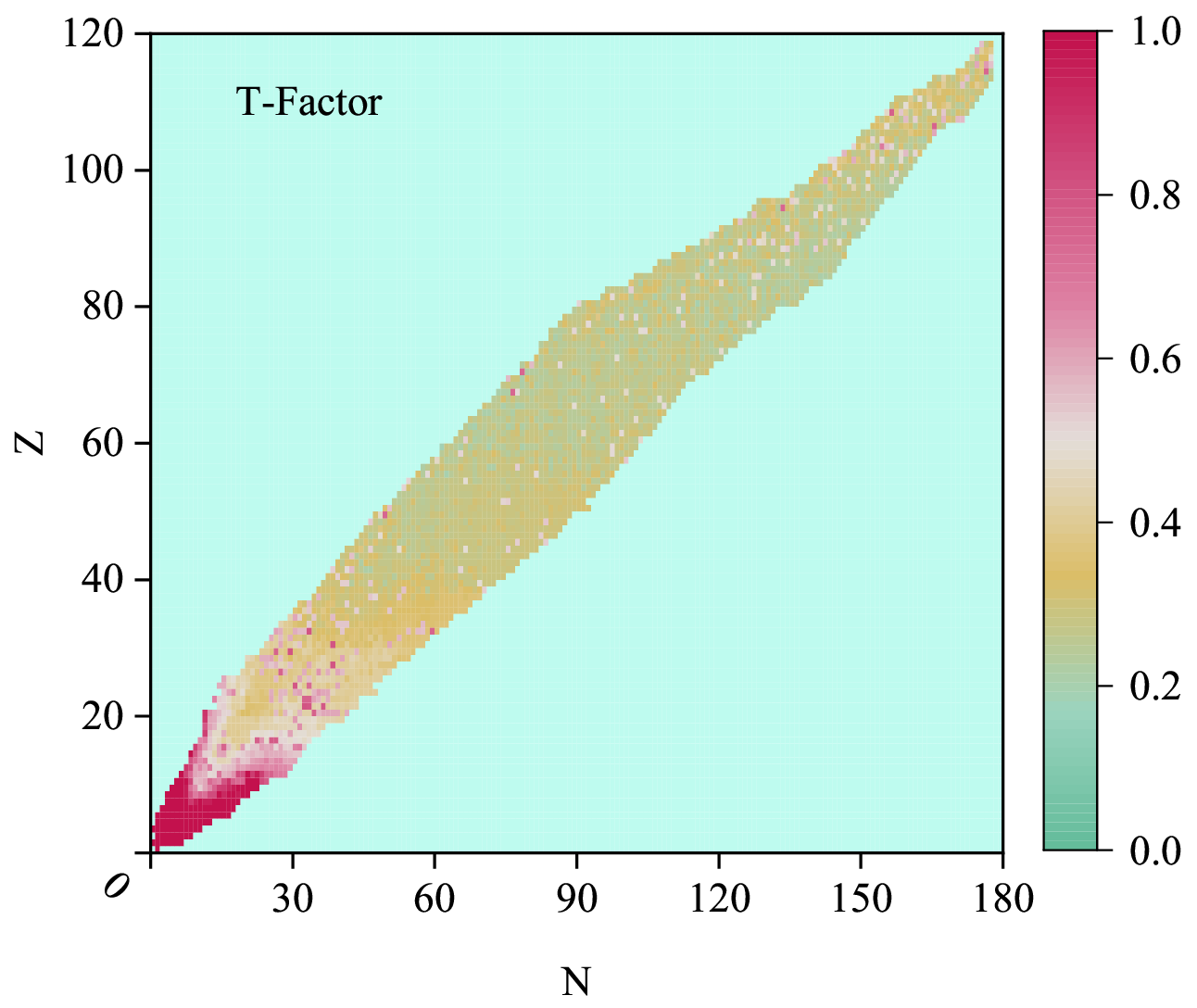}
\caption{(Color online) The $T$-factor for nuclides represents the relative thickness of
the surface liquid-like layer. A higher $T$-factor indicates that the surface high-density region occupies a larger proportion of the radius.}

\label{fig:T_Factor}
\end{figure}

\begin{figure}[!htb]
\includegraphics[width=\hsize] {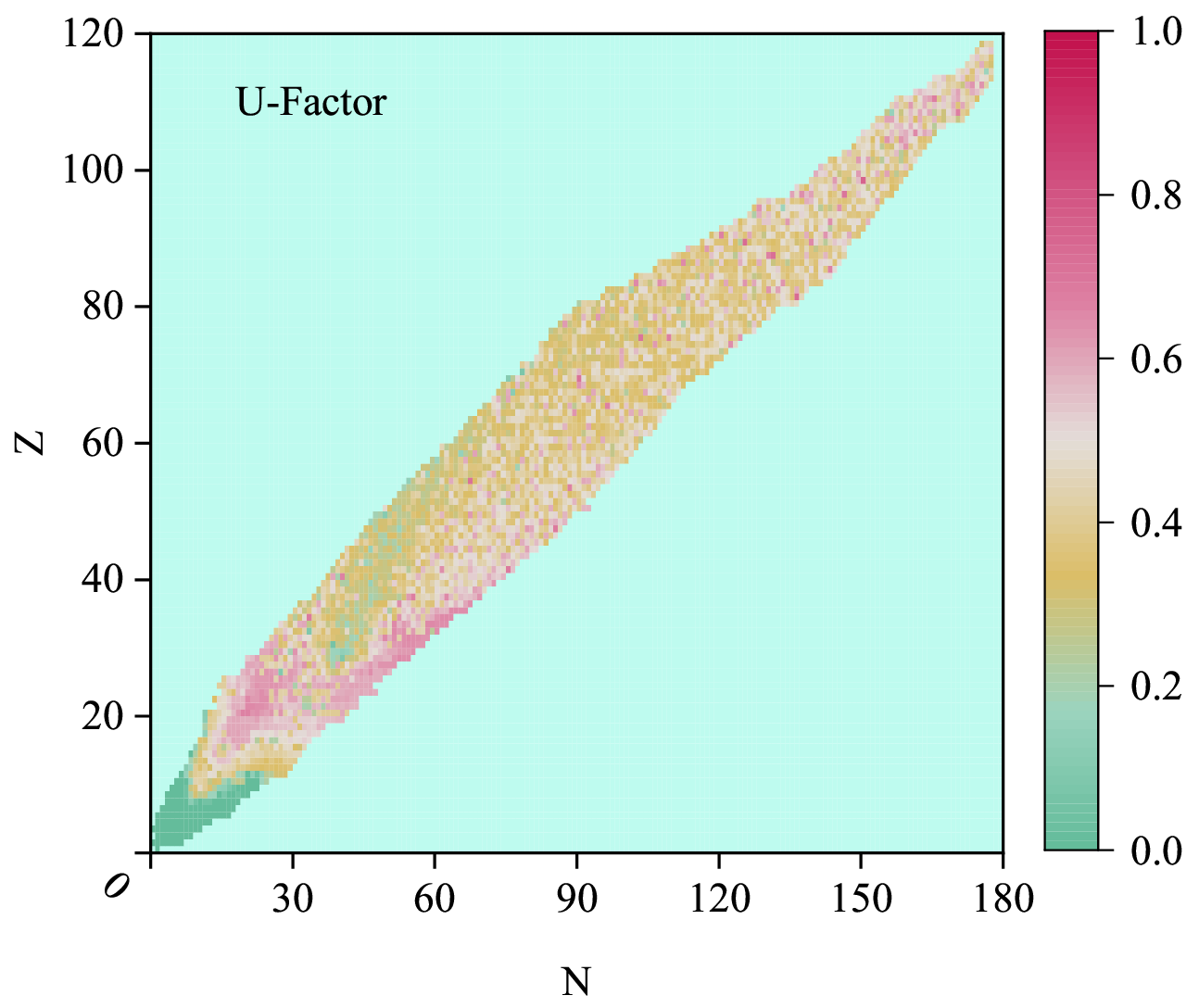}
\caption{(Color online) The $U$-factor for nuclides represents the relative size of the bubble region. A higher $U$-factor indicates a larger proportion of bubble radius in the bubble and toroidal bubble nuclei.}

\label{fig:U_Factor}
\end{figure}

Since both $T$ and $U$ are dimensionless quantities and are computed in a similar manner, Figs.~\ref{fig:T_Factor} and \ref{fig:U_Factor} employ the same numerical range and color map. The colors in the figures allow clear differentiation of the relative contributions of the surface liquid-like layer and the internal bubble size to the overall nuclear structure across different types of nuclides. From the overall color distributions in both figures, the patterns still broadly follow the classification shown in Fig~\ref{fig:B_Factor}. Light-mass droplet nuclei exhibit $T=1$ and $U=0$, while the transition region between droplet and bubble configurations gradually evolves toward $T\approx0.5$ and $U\approx0.4$. In the vicinity of $^{40}$Ca as well as the intermediate-mass region with high neutron excess, where the $H$factor, representing the degree of central density depletion,is relatively large, values of $T\approx0.35$ and $U\approx0.65$ are observed. The distribution of $T$ and $U$ shows a clear boundary between these two regions that closely parallels the pattern of the $H$-factor. As the mass number increases and the system enters the region dominated by toroidal bubble structures, the color representing the $T$-factor transitions into a yellow-green mixture. This reflects the fact that with increasing radius, the spatial extent of surface-cluster wave packets becomes progressively smaller relative to the total size, leading to a reduced proportion of surface thickness. Certain nuclei near the drip lines exhibit anomalously high $T$ values as a result of their inherent instability and enhanced surface fluctuations. Meanwhile, the $U$-factor remains predominantly concentrated in the range 0.4-0.6, indicating that the relative size of the bubble region is relatively stable in this domain.

For light nuclei, the structures are intrinsically anisotropic and cannot be described as isotropic spheres. However, as the $B$-factor indicates, there is no low-density region at the center with $H=0$, $T=1$, and $U=0$, meaning that cluster structure does not alter the parameter values. For medium-mass nuclei, such as $^{28}\mathrm{Si}$ shown in Fig.~\ref{fig:density_distribution}~(b), the wave packet dispersion is sufficient to form a fully surrounded bubble, so the $B$-factor classification is not affected, but the values of $H$, $T$, and $U$ may differ from those of a uniform spherical distribution. For heavier nuclei, the nucleon distribution itself tends toward sphericity, and wave packets tend to overlap significantly, resulting in smoother and more pronounced isotropy. In these systems, radial classification can almost directly reflect the actual reduction in density, with minimal limitations regarding anisotropy.

Overall, this paper primarily focuses on establishing a classification-quantification framework for bubble-like structures, enabling the quantitative scaling of their density distribution characteristics for subsequent studies of such exotic nuclear structures. While the global scan of the nuclide chart is influenced by the specific model assumptions and cluster-structure characteristics, the present results should be regarded as a useful reference for future studies.

\section{Summary}

This study presents a systematic investigation of low-energy cluster configurations across all known nuclides in the AME2020 database, based on the friction cooling mechanism implemented in the extended quantum molecular dynamics (EQMD) model. The primary objective is to explore the emergence of bubble-like structures in nuclear density distributions and to establish a unified classification-quantification framework for their classification and characterization. This methodology critically relies on the generation of relaxed low-energy cluster states, where intrinsic cluster formation significantly modulates local density distributions to produce bubble-like structures unique to cluster-dominated configurations. The friction cooling in EQMD is essential for stabilizing these states, making it possible to study structural features beyond the scope of static mean-field approaches.

A structural classification factor $B$ is introduced by analyzing inflection points in radial density profiles, which categorizes nuclei into three types: droplet ($B=0$), bubble ($B=1$), and toroidal bubble ($B=2$). To quantify the degree of central density depletion, an additional parameter $H$ is defined, complemented by two dimensionless quantities $T$, representing the relative thickness of the surface, and $U$, characterizing the relative size of the internal bubble region. Together, these form a multidimensional quantitative descriptor that not only distinguishes between different density characteristics of bubble-like nuclei but also reveals systematic trends in nuclear structure as functions of mass number and isospin asymmetry.

The results indicate that droplet configurations dominate in light nuclei, bubble structures are prevalent in the medium-mass region, and toroidal bubble configurations begin to appear at proton numbers $Z\approx25$, becoming increasingly common in heavy nuclei. In particular, the vicinity of $^{40}$Ca and the neutron-rich medium-mass region, consistent with previous studies, exhibit the highest $H$ values alongside significant $U$ values, identifying both regions as prime candidates for experimental searches for bubble structures. Furthermore, the widespread occurrence of bubble structures in the superheavy region is also consistent with earlier theoretical predictions.

In summary, this work overcomes the limitation of traditional mean-field theories, which rely on smooth, homogeneous density distributions, by capturing cluster configurations through the EQMD framework. The focus on relaxed low-energy cluster states unveils a rich landscape of nuclear shapes and enables a systematic classification across the nuclear chart. The proposed $BHTU$ parametric unified classification-quantification framework provides a robust theoretical tool for the identification and prediction of bubble-like nuclei, and provides guidance for future experiments at radioactive ion beam facilities. These findings deepen our understanding of nuclear matter and lay the foundation for exploring the complexity of the nuclear density  distribution.

\appendix
\setcounter{figure}{0}
\renewcommand{\thefigure}{A\arabic{figure}}
\section{Proton and neutron density characteristics}
In previous theoretical and experimental studies, the identification of nuclear bubble structures has frequently relied on specific density distributions arising from the absence of $s$ orbitals for protons or neutrons within the static mean-field picture, necessitating separate depletion maps to verify these specific scenarios. However, bubble-like central depletion discussed in this work originates primarily from local density non-uniformities induced by an alpha-dominated cluster structure characterized by neutron-proton symmetry pairing. Even within these specialized relaxed low-energy cluster states, the independent analysis of proton and neutron densities remains highly significant. Such analysis not only permits the direct verification of the proposed structure but also enhances the understanding of cluster types, distributions, and nuclear molecular configurations. Consequently, the figure regarding the characteristics of proton and neutron density distributions has been added in the appendix. As shown in Fig.~\ref{fig:p_n}, for medium-mass nuclei, the proton bubbles are primarily distributed in symmetric nuclei and regions with a slight excess of neutrons. This distribution arises from the isospin-symmetric pairing of neutrons and protons within alpha clusters, combined with central repulsion associated with neutron excess and the surface delocalization of neutrons near the drip line. In contrast, the neutron bubbles exhibit a distribution on both sides of the chart for the same reason.

\begin{figure}[!htb]
\includegraphics[width=\hsize] {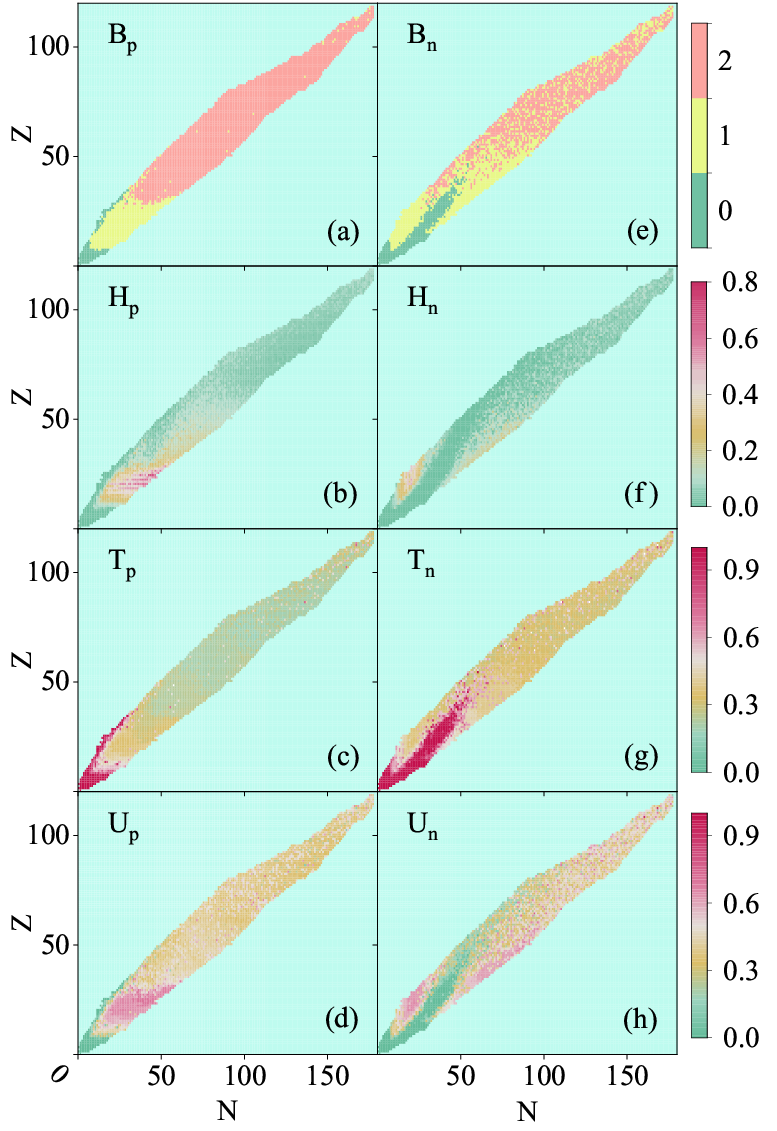}
\caption{(Color online) The $BHTU$-factor for proton (left) and neutron (right) of nuclides.}
\label{fig:p_n}
\end{figure}

\end{document}